# General theory of emergent elasticity for second-order topological phase transitions

*Yangfan Hu\*,*

*Research Institute of Interdisciplinary Science & School of Materials Science and Engineering, Dongguan University of Technology, Dongguan, 523808, China.*

**Abstract**

The raise of the symmetry breaking mechanism by Landau[1] is a landmark in the studies of phase transitions. The Kosterlitz-Thouless phase transition[2-3] and the fractional quantum Hall effect[4], however, are believed to be induced by another mechanism: topology change. Despite rapid development of the theory of topological orders[5-7], a unified theoretical framework describing this new paradigm of phase transition and its relation to the Landau paradigm, is not seen. Here, we establish such a framework based on variational principle, and show that the critical condition for any second-order topological phase transitions is loss of positive-definiteness of a topologically protected second-order variation of the free energy. A topologically protected variation of a field solution is performed with respect to its "emergent displacements", which is introduced by constructing an emergent elasticity problem for the solution. The Landau paradigm of phase transitions studies global property changes induced instability, while topological phase transitions study local property changes induced topological instability. The general effectiveness of this criterion is shown through analyzing the topological stability of several prototype solutions in both real space (two-kink solutions of the sine-Gordon model; an isolated skyrmion in chiral magnets) and reciprocal space (phonon spectrum of a monoatomic chain and a diatomic chain; band structure of a monoatomic chain within the Kronig-Penny model). Every field is emergent elastic with spatially modulated stiffness, and changes of topological property occur at its softened points. We anticipate our work to be a starting point for a general study of topological phase transitions in all field theories.

\* Corresponding author: huyf@dgut.edu.cn

**Main**

Based on the symmetry breaking mechanism[1], Landau uses group theory to simplify the free energy functional minima problem, which reduces the Euler-Lagrangian equation to a set of algebraic equations of the order parameters, greatly facilitating its solving. This theory of symmetry breaking mediated phase transition (SBPT) successfully explains a lot of phase transitions not only between equilibrium states[8-9], but also of some non-equilibrium processes[10-12]. However, since the Landau paradigm is designed to study changes of bulk properties, it is intrinsically inapplicable to any phase transitions induced by local property changes such as the Kosterlitz-Thouless (KT) phase transition[2-3] and the Fractional quantum Hall (FQH) effect[4]. In its original definition[1], the order parameters are combination of coefficients of the irreducible representation of the state variables, for which change of value of the order parameters means change of the free energy density. When some local field pattern of the state variables changes, the related bulk free energy change corresponds to a finite value, and hence the corresponding free energy density change is zero. The KT phase transition describes detaching of a pair of topological defects at some critical temperature, and the associated energy change is finite. The FQH effect also describes a change of local property, not in the real space but in the reciprocal space.

A possible way to study local property change induced phase transitions is to perform stability analysis for solutions of the Euler-Lagrangian equation. This is usually conducted by considering the positive-definiteness of the second-order variation of the free energy functional[13]. However, to study topology change mediated phase transitions (TPT), the concept of topology must be incorporated. A topologically protected change of a field solution means that a continuous transformation links the field pattern before and after the change. And the mathematical language to describe this continuous transformation has been well established in continuum mechanics.

By introducing a new concept "emergent displacement" from continuum mechanics, we establish the emergent elasticity problem for an arbitrary solution of any static Euler-Lagrangian equation. We prove that a stable solution of the Euler-Lagrangian equation corresponds to zero solution of the emergent displacements, and the second-order variation with respect to the emergent displacements determines the topological stability of the solution. To be more specific, loss of positive-definiteness of the second-order variation, or the "emergent stiffness matrices", at any spatial point provides a necessary condition for the occurrence of a TPT. On this basis, we are able to clarify the difference between the SBPT and the TPT: (a) occurrence of a TPT means the topological property of the solution changes, but the solution may still be stable, so TPT describe phase transitions that are more subtle than SBPT; (b) SBPT study changes of bulk properties while TPT study changes of local properties. For 1D scalar fields, We derive the general form of the critical condition of topological stability for any 1D scalar fields, and find that emergent stiffness of a field solution stems from a nontrivial topological density. We further apply the theory to study the topological stability of two prototype models in real space (the two-kink solutions of the sine-Gordon model[14-16], an isolated skyrmion with fixed modulus in chiral magnets[17, 18]) and three prototype models in the reciprocal space (the phonon spectrum of a monoatomic chain, the band gap opening of the phonon spectrum of a diatomic chain by changing the mass ratio, and the electronic band sturcture of a 1D crystal with the Kronig-Penney (KP) model[19, 20]). In all cases studied, topology change occurs at those points in the real or reciprocal space with a vanishing emergent stiffness.

**Mathematical foundation of emergent elasticity and topological stability**

*A. What is an emergent displacement ?*

Consider a 1D scalar field $y(x)$ undergoing a topologically protected change in the domain, then the field solution after this change can generally be written as $y(x - u(x))$. In other words, the function form of $y$ does not change, but the domain where it is defined in undergoes some "displacement" $u(x)$. Because this displacement does not really occur in space, we call $u(x)$ an emergent displacement field, in order to be distinguished from the real displacement field studied in solid mechanics. As an example, we compare the function image of $y(x) = x, u(x) = 0$ (Figure 1(a)) and $y(x) = x, u(x) = \sin x$ (Figure 1(c)), and the effect of $u(x)$ is presented by the inhomogeneous distribution of vertical geodesics in Figure 1(c). If the space maintains a continuum after the emergent displacement, we have $u'(x) \in (-\infty, 1) \ \forall \ x$. In the example $u(x) = \sin x$, we have $u'(x) = 1$ at $x = n\pi, (n = 0, \pm 1, \pm 2, \dots)$. It is shown in Figure 1(c) that at these points the relative elongation locally reaches infinity (i.e., at these points the domain is locally teared apart).

Similarly, we can also define the emergent displacement field $v(y)$ in the range, that is $y(x) \to$

$(y + v(y))(x)$ (Figure 1(d)), and a more general case where both the domain and the range undergo emergent displacements, that is $y(x) \to (y + v(y))(x - u(x))$ (Figure 1(b)).

From the example we can extract the following general rules concerning emergent displacement:

1. The general expression of a topologically protected change of a 1D scalar field $y(x)$ is $y(x) \to (y + v(y))(x - u(x))$, where $u'(x) \in (-\infty, 1)$ in the domain and $\frac{dv}{dy}(y) \in (-1, \infty)$ in the range. A topologically protected change of $y(x)$ does not change its monotonicity at any point.

2. For $y(x)$ with unbounded domain and range, a topologically protected change of $y(x)$ does not correspond to a unique choice of $u(x)$ and $v(y)$. For example, the function images illustrated in Figure 1(c) and Figure 1(d) are the same, but they come from different choices of $u(x)$ and $v(y)$.

3. When a small change of $y(x)$ can be described by an emergent displacement field of the underlying domain or range, the related energy change is the corresponding emergent elastic energy. <u>For an emergent displacement beyond topological protection to occur continuously at any point, the related local emergent stiffness coefficient should vanish.</u>

B. *Topologically protected variation and emergent elasticity of a 1D scalar field*

We first consider a free energy minimization problem where the free energy is described by

$$\Phi = \int \phi(x, y, y') \, dx, \tag{1}$$

with $\phi(x, y, y')$ the free energy density and the state variable is a 1D scalar field $y(x)$. Minimization of $\Phi$ with respect to $y(x)$ yields the Euler-Lagrangian equation

$$\frac{d}{dx}\left(\frac{\partial \phi}{\partial y'}\right) - \frac{\partial \phi}{\partial y} = 0. \tag{2}$$

Assume that $y = y_0(x)$ minimizes the free energy defined by eq. (1), that means $y_0$ satisfies eq. (2), and $\delta^2 \Phi$ must be larger than zero[13], which yields (I) $\left(\frac{\partial^2 \phi}{\partial y'^2}\right)_0 > 0$ or (II) $\left(\frac{\partial^2 \phi}{\partial y'^2}\right)_0 = 0$, $\left[\frac{\partial^2 \phi}{\partial y^2} - \frac{d}{dx}\left(\frac{\partial^2 \phi}{\partial y' \partial y}\right)\right]_0 > 0$ for any $x$. The requirement of $\left(\frac{\partial^2 \phi}{\partial y'^2}\right)_0 = 0$ in (II) usually means that $\phi$ is irrelevant to $y'$, in which case we can rewrite (II) as $\phi = \phi(x, y)$ and $\left(\frac{\partial^2 \phi}{\partial y^2}\right)_0 > 0$ for any $x$. Here a subscript 0 means the term takes value at $y = y_0(x)$.

Now we construct the domain emergent elasticity problem of the solution $y = y_0(x) \to y_0(x - $

$u(x))$, whose emergent elastic energy density is denoted by $\phi_u = \phi(x, y_0(x-u(x)), \frac{d}{dx} y_0(x-u(x)))$. Similarly, we can define the range emergent elasticity problem of the solution $y = y_0(x) \to y_0(x) + v(y_0(x))$, whose emergent elastic energy density is denoted by $\phi_v = \phi(x, y_0(x) + v(y_0(x)), \frac{d}{dx}[y_0(x) + v(y_0(x))])$. Here $u(x)$ and $v(y)$ takes small values at everywhere they are defined. Since $y = y_0(x)$ is a stable solution of $\frac{\delta \Phi}{\delta y} = 0$, it can be proved that (Methods A) the two types of emergent elasticity problems $\frac{\delta \Phi_u}{\delta u} = 0$ and $\frac{\delta \Phi_v}{\delta v} = 0$ correspond to zero solutions $u(x) = 0$ and $v(y_0(x)) = 0$. Consider the two related replacements (a) $y_0(x) \to y_0(x - u(x))$ and (b) $y_0(x) \to y_0(x) + v(y_0(x))$, $\delta u$ and $\delta v$ can on one hand be regarded as regular variation of the vanishing solution $u(x) = 0$ and $v(y_0(x)) = 0$, but on the other hand they can also be regarded as restricted variation of $y$ at $y = y_0(x)$, for which we call them topologically protected variation of the stable solution $y = y_0(x)$. The stability of the vanishing solution $u(x) = 0$ and $v(y_0(x)) = 0$ are determined by the positiveness of the second-order variation of $\Phi_u$ and $\Phi_v$ with respect to $u$ and $v$[13], for which we have

$$\delta^2 \phi_u = \frac{1}{2}(C^\varepsilon \delta u'^2 + C^u \delta u^2), \tag{3}$$

$$\delta^2 \phi_v = \frac{1}{2}(C^w \delta v^{\circ 2} + C^v \delta v^2), \tag{4}$$

where $C^\varepsilon = \left(\frac{\partial^2 \phi}{\partial y'^2} y'^2\right)_0$ is called domain emergent elastic stiffness constant, and $C^u = \left(\frac{\partial^2 \phi}{\partial y^2} y'^2 - \frac{d}{dx}(\frac{\partial^2 \phi}{\partial y'^2}) y'y'' - \frac{\partial^2 \phi}{\partial y'^2} y'y''' - \frac{d}{dx}(\frac{\partial^2 \phi}{\partial y \partial y'}) y'^2\right)_0$ is called domain emergent displacement stiffness constant. $C^w = \left(\frac{\partial^2 \phi}{\partial y'^2} y'^2\right)_0$ is called range emergent elastic stiffness constant, $C^v = \left(\frac{\partial^2 \phi}{\partial y^2} - \frac{d}{dx}(\frac{\partial^2 \phi}{\partial y \partial y'})\right)_0$ is called range emergent displacement stiffness constant, and $v^\circ = (\frac{dv(y)}{dy})_0$. A general extension of the theory developed above to dynamical problems is difficult, due to a well-known fact that generally the stationary action principle does not correspond to a functional minimization problem[21]. Nevertheless, we will see later that in some special cases where a minimization potential can be found, the theory is still applicable.

C. *Critical condition of topological stability of field solutions*

The critical condition of topological stability of the 1D scalar field solution $y = y_0(x)$ can immediately be derived as follow. When $\phi$ is relevant to $y'$ (case (I)), it means there exists a nonzero choice of $\delta u'$ and $\delta u$ such that $\delta^2 \phi_u$ defined in eq. (3) is zero at some point or points in space for domain emergent elasticity, and it means there exists a nonzero choice of $\delta v^o$ and $\delta v$ such that $\delta^2 \phi_v$ defined in eq. (4) is zero at some point or points in space for range emergent elasticity. According to variational principles[13], the $\frac{1}{2} C^\varepsilon \delta u'^2$ term dominates $\delta^2 \phi_u$ in eq. (3) and the $\frac{1}{2} C^w \delta v^{o2}$ term dominates $\delta^2 \phi_v$ in eq. (4), for which the critical condition in this case is $C^\varepsilon = 0$ for domain emergent elasticity and $C^w = 0$ for range emergent elasticity. When $\phi$ is irrelevant to $y'$ (case (II)), the critical condition in this case is $C^u = 0$ for domain emergent elasticity and $C^v = 0$ for range emergent elasticity.

The critical condition can be further specified since analysis of topological stability is performed under the premise that $y = y_0(x)$ minimizes the free energy, which is fulfilled only under the following two cases: (I) $\left(\frac{\partial^2 \phi}{\partial y'^2}\right)_0 > 0$ or (II) $\phi = \phi(x, y)$ $and$ $\left(\frac{\partial^2 \phi}{\partial y^2}\right)_0 > 0$. For a 1D scalar field, the critical condition of topological stability under different cases is listed in Table 1. We have also studied a more generalized form of emergent displacement $y(x) \rightarrow (y + v(y))(x - u(x))$. Due to the fact that a topologically protected change of $y(x)$ does not correspond to a unique choice of $u(x)$ and $v(y)$, we find that no further useful information of the critical condition of topological stability can be obtained.

Table 1. Critical condition of topological stability for a 1D scalar field under different conditions.

|  | (I) $\left(\frac{\partial^2 \phi}{\partial y'^2}\right)_0 > 0$ | (II) $\phi = \phi(x, y)$ $and$ $\left(\frac{\partial^2 \phi}{\partial y^2}\right)_0 > 0$ |
| --- | --- | --- |
| Domain (Case (a)) | $y' = 0$ | N/A |
| Range (Case (b)) | $y' = 0$ | $y' = 0$ |

From Table 1, the condition of topological stability for all available cases is: y' = 0, which shows that the origin of emergent stiffness is spatial variation of the field. Since for 1D scalar field y' happen to be the topological density [14], we have in this specific case: emergent stiffness stems from topological density.

For an arbitrary vector solution $\mathbf{p}(\mathbf{x}) = \tilde{\mathbf{p}}(\mathbf{x})$ with $n$ components in $m$-dimensional space, we extended the derivation above and give a general form of critical condition of topological stability.

For case (a), we rewrite $\widetilde{\mathbf{p}}(\mathbf{x}) \to \widetilde{\mathbf{p}}(\mathbf{x} - \mathbf{u}(\mathbf{x}))$, where $\frac{d\mathbf{u}}{d\mathbf{x}}$ corresponds to a $m \times m$ matrix. Define the domain emergent elastic stiffness matrix $\mathbf{C}^{\boldsymbol{\varepsilon}}$ and domain emergent displacement stiffness matrix $\mathbf{C}^{\mathbf{u}}$, whose components are

$$C_{ij}^{\varepsilon} = \frac{\partial^2(\delta^2 \phi_u)}{\partial(\delta \varepsilon_i) \partial(\delta \varepsilon_j)}, \tag{7}$$

and

$$C_{ij}^{u} = \frac{\partial^2(\delta^2 \phi_u)}{\partial(\delta u_i) \partial(\delta u_j)}. \tag{8}$$

Here $\boldsymbol{\varepsilon}$ is a vector whose components are all the components of the matrix $\frac{d\mathbf{u}}{d\mathbf{x}}$. When $\phi$ is relevant to the gradients of $\mathbf{p}(\mathbf{x})$, the critical condition of topological stability is at some point or points in space, at least one of the eigenvalues of $\mathbf{C}^{\boldsymbol{\varepsilon}}$ drops to zero. When $\phi$ is irrelevant to the derivatives of $\mathbf{p}(\mathbf{x})$, the critical condition of topological stability is at some point or points in space, at least one of the eigenvalues of $\mathbf{C}^{\mathbf{u}}$ drops to zero.

Similarly, for case (b) we rewrite $\widetilde{\mathbf{p}}(\mathbf{x}) \to \widetilde{\mathbf{p}}(\mathbf{x}) + \mathbf{v}(\widetilde{\mathbf{p}}(\mathbf{x}))$, and $\mathbf{v}° = \frac{d\mathbf{v}}{d\mathbf{p}}$ corresponds to a $n \times m$ matrix. Define the range emergent elastic stiffness matrix $\mathbf{C}^{\mathbf{w}}$ and range emergent displacement stiffness matrix $\mathbf{C}^{\mathbf{v}}$, whose components are

$$C_{ij}^{w} = \frac{\partial^2(\delta^2 \phi_v)}{\partial(\delta w_i) \partial(\delta w_j)}, \tag{9}$$

and

$$C_{ij}^{v} = \frac{\partial^2(\delta^2 \phi_v)}{\partial(\delta v_i) \partial(\delta v_j)}. \tag{10}$$

Here $\mathbf{w}$ is a vector whose components are all the components of the matrix $\mathbf{v}°$. When $\phi$ is relevant to the gradients of $\mathbf{p}(\mathbf{x})$, the critical condition of topological stability is at some point or points in space, at least one of the eigenvalues of $\mathbf{C}^{\mathbf{w}}$ drops to zero. When $\phi$ is irrelevant to the derivatives of $\mathbf{p}(\mathbf{x})$, the critical condition of topological stability is at some point or points in space, at least one of the eigenvalues of $\mathbf{C}^{\mathbf{v}}$ drops to zero.

We further prove in Methods B that when $n = m$ and when $\phi$ takes some regular form, the critical condition of topological stability obtained from analysis of domain emergent elasticity are equivalent to that obtained from analysis of range emergent elasticity.

Generally, the critical condition of topological stability is a necessary but not sufficient condition for the occurrence of a second-order TPT. To explain why, consider the following example: if one of the eigenvalues of $\mathbf{C}^\varepsilon$ reaches zero and the critical condition of topological stability is attained, the eigenvector corresponding to the vanishing eigenvalue of $\mathbf{C}^\varepsilon$ determines the particular form of disturbance that will be unstable. However, if this type of disturbance is not allowed by the external field under consideration, then a TPT will not occur. An appropriate example is shown below when studying the TPT of an isolated skyrmion in helimagnets. Moreover, the criterion developed above is only valid for a second-order TPT. We find that the annihilation of an isolated skyrmion with changeable modulus corresponds to a first-order TPT[22], where at the phase transition point (center of a skyrmion) the emergent stiffness is not softened according to static analysis. However, further dynamical soft mode analysis show that combining the equilibrium field configuration and the soft mode vibration, the center point of a skyrmion is dynamically softened which we believe initiates a first-order TPT[22].

### D. Relation between TPT and SBPT

The occurrence of a SBPT is determined by the positive-definiteness of the second-order variation condition, where the calculation is greatly simplified by using group theory. The occurrence of a TPT, on the other hand, is determined by the positive-definiteness of the topologically protected second-order variation, which is calculated based on the assumption that the solution has a positive second-order variation. In this sense, TPT is more subtle than SBPT.

The Landau paradigm of phase transition only concerns about the symmetry before and after the phase transition, and hence the free energy density constructed is independent of the solution of state variables, which shows its universality. The new paradigm developed for TPT construct an emergent elasticity problem for any solution of the state variables under consideration, and discuss the topological stability of the solution within a unified mathematical framework of emergent elasticity. In this sense, the new paradigm retains the universality of the Landau paradigm.

SBPT describe changes of macroscopic properties while TPT describe changes of local properties. The change of topological property of a solution is always initiated at one or some local points in space, so that the emergent stiffness is intrinsically a spatial-dependent local property instead of a macroscopic property, extending the traditional understanding of stiffness in solid mechanics.

The significance of the order parameter is different for SBPT and TPT. For the Landau paradigm dealing with SBPT, the choice of order parameters is vital since the free energy is expressed by a function of the order parameters, and solution and analysis of the order parameters are all one can do to understand a SBPT. On the other hand, we can also define appropriate order parameters for a TPT[5, 6, 23], but generally they only appear as part of the unknowns of the system, so they usually appear only in the last part of the analysis to help understand the physics underneath the complicated solution.

E. *Relation between ordinary theories of topology and the theory of emergent elasticity and topological stability*

The concept of topology can be constructed without introducing the notion of "distance", or a metric[19]. In many of the applications of topology in physics, what matters is "continuity minus metric". In this sense, we categorize different field solutions in terms of their topological properties and say that solutions with different topological charge cannot be transformed to each other and this is a property called "topological protection". However, topological protection provides conditional stability instead of permanent stability for a solution, otherwise no particle can appear from the vacuum because they have different topological charge. The theory of emergent elasticity and topological stability aims at finding out the critical condition under which solutions with different topological charge can transform to each other, and the very first concept introduced is the emergent displacement field, or a metric.

**Emergent elasticity and topological stability in real space**

A. *Stability of the two-kink solutions of the sine-Gordon model*

Consider the two-kink solutions

$$y(x,t) = s_1 = 4\tan^{-1}\left(\frac{\sinh x}{\cosh t}\right), \tag{11}$$

$$y(x,t) = s_2 = -4\tan^{-1}\left(\frac{\sinh t}{\cosh x}\right) \tag{12}$$

of the 1D sine-Gordon model[14-16] with a Lagrangian density

$$L = \frac{1}{2}\dot{y}^2 - \frac{1}{2}y'^2 - (1-\cos y), \tag{13}$$

where $\dot{y} = \frac{\partial y}{\partial t}$, $y' = \frac{\partial y}{\partial x}$ and the potential energy density $\phi = \frac{1}{2}y'^2 + (1-\cos y)$.

In Eqs. (11, 12), all related parameters are set to unity for simplicity. Eq. (11) describes the elastic collision of two kinks with a repulsive interaction (Fig. 2(a)), while Eq. (12) describes the elastic collision of a kink-antikink pair with an attractive interaction (Fig. 3(a)). For these two 1D soliton solutions, only one effective emergent stiffness coefficient exists:

$$C^e = y'^2, \qquad (14)$$

which represents both the domain and range emergent stiffness coefficients. For the two solutions given in Eqs. (11, 12), we have

$$C^{e1} = C^e(s_1) = \frac{16 \cosh^2 x \ \text{sech}^2 t}{(1+\text{sech}^2 t \sinh^2 x)^2}, \qquad (15)$$

$$C^{e2} = C^e(s_2) = \frac{64 \sinh^2 x \ \sinh^2 t}{(\cosh 2t + \cosh 2x)^2}. \qquad (16)$$

Fig. 2(b) and Fig. 3(b) show the distributions of $C^{e1}$ and $C^{e2}$ at different times. In the kink-kink collision described by Eq. (11), the emergent stiffness at the collision point $C^{e1}(x = 0, t)$ gradually increases as the two kinks approach each other and finally reaches its maximum at $t = 0$ when the two kinks collide (Fig. 2(b)). This phenomenon resembles the collision of two elastic balls, which squeeze and stabilize each other after contact and harden the contact area.

In the kink-antikink collision described by Eq. (12), the kink and antikink scatter elastically and annihilate at $t = 0$ (Fig. 2(a)). Interestingly, according to Eq. (16), $C^{e2}(x = 0, t) = 0$ for all $t$ (Fig. 3(b)), forming a topological instability point for both the kink and antikink. Regarding $t$ as a parameter, we can describe the kink-antikink collision process by a topological phase transition, which is initiated at the contact point as well as the topological instability point $x = 0$. To see this, we calculate the topological charge of the left half-space

$$N_{lh} = \frac{1}{2\pi}\int_{-\infty}^{0} s_2' dx = \frac{1}{2\pi}s_2(x = 0) = \frac{2}{\pi}\tan^{-1}(\sinh t), \qquad (17)$$

which, at $t < 0$, represents the topological charge of the antikink occupying the left half-space. As shown in Fig. 3(c), this topological charge is not conserved and continuously varies with time. At $t = 0$, $N_{lh}$ reaches zero and increases as $t$ increases, which means that the kink and the antikink have moved pass each other at $t = 0$, and for $t > 0$, the kink occupies the left half-space.

### B. Stability of an isolated skyrmion in chiral magnets under an applied magnetic field

We then consider the three-dimensional magnetization vector field in B20 chiral magnets at low temperatures, which permits a static soliton solution called an isolated magnetic skyrmion of the Bloch type[17, 18] (Fig. 4(a)). The potential energy density of the system is given in a rescaled form[24] as

$$\phi = \sum_{i=1}^{3}\left(\frac{\partial \mathbf{m}}{\partial x_i}\right)^2 - 2\mathbf{b}\cdot\mathbf{m} + \mathbf{m}\cdot(\nabla\times\mathbf{m}), \tag{18}$$

where the vector field $\mathbf{m}$ denotes the magnetization, whose length is restricted to be 1 in this model. $\mathbf{b}$ denotes the external magnetic field vector, and we assume here that $\mathbf{b} = [0, 0, b]^T$. The functional given in Eq. (18) can also be applied to describe skyrmions in liquid crystals[25]. We describe the problem in cylindrical coordinates $\mathbf{x} = [\rho\cos\varphi, \rho\sin\varphi, z]^T$, and the isolated skyrmion solution takes the form

$$\mathbf{m} = [-\sin\theta(\rho)\sin\varphi, \sin\theta(\rho)\cos\varphi, \cos\theta]^T, \tag{19}$$

where $\theta(\rho)$ is to be numerically determined through minimization of $\int \phi\, dV$ with the boundary conditions $\theta(0) = \pi$ and $\theta(R) = 0$, $R$ being the radius of an isolated skyrmion. In Fig. 4(c), the profile of $\theta(\rho)$ solved at different values of $b$ is shown. An isolated skyrmion is a two-dimensional symmetrical structure of magnetization with radius $R$, whose dependence on $b$ is shown in Fig. 4(b). According to the theory of emergent elasticity and topological stability for 3D vector fields (see Methods C for details), the emergent stiffness matrix for the isolated skyrmion solution given in Eq. (19) can be diagonalized in polar coordinates, and we have four diagonal elements: $C^e_{\rho\rho} = 2\theta'^2(\rho)$, $C^e_{\varphi\varphi} = \frac{2\sin^2\theta(\rho)}{\rho^2}$, $C^e_{\gamma\gamma} = 2\left(\theta'^2(\rho) + \frac{\sin^2\theta(\rho)}{\rho^2}\right)$, and $C^{\omega_3} = 2\left(\theta'^2(\rho) + \frac{\sin^2\theta(\rho)}{\rho^2}\right)$. The positive definiteness of these four coefficients at every spatial point occupied by an isolated skyrmion guarantees its topological stability. In contrast, when any one of these four parameters vanishes at any spatial point, the topological stability of an isolated skyrmion is broken, and the corresponding deformation modes are as follows: $C^e_{\rho\rho}$ and $C^e_{\varphi\varphi}$ correspond to emergent elongation in the $\rho$ and $\varphi$ directions, respectively, in the $\rho - \varphi$ plane, $C^e_{\gamma\gamma}$ corresponds to emergent shearing in the $\rho - \varphi$ plane, and $C^{\omega_3}$ corresponds to emergent rotation in the $\rho - \varphi$ plane. The distribution of these four emergent stiffness coefficients as functions of $\rho$ at different values of $b$ are plotted in Fig. 4(d-f).

Due to the external boundary condition $\theta(R) = 0$, we always have $C^e_{\varphi\varphi}(\rho = R) = 0$, which means that an isolated skyrmion is intrinsically topologically unstable with respect to any disturbance that breaks the axial symmetry at $\rho = R$. In contrast, the topological stability of a skyrmion with respect to an axially symmetric disturbance (e.g., a change in a homogeneous magnetic field $b$ along the $z$ axis) is determined by $C^e_{\rho\rho}$. A well-known fact is that as $b$ increases to some critical value $b_c$, $R$ approaches infinity[17] (Fig. 4(b)), characterizing a transition from a localized state to a state that occupies the whole space. As shown in Fig. 4(b), this transition is induced by softening of $C^e_{\rho\rho}(\rho = R)$ at $R \to \infty$. This phenomenon can be regarded as the single particle version of the Kosterlitz-Thouless phase transition[2-3], so we call it a single-particle Kosterlitz-Thouless (SPKT) phase transition. It is a topological phase transition because the form of the solution does not change or become unstable, but we can find an appropriate order parameter: the averaged topological density $\rho_t = \frac{1}{4\pi^2 R^2} \int \mathbf{m} \cdot (\mathbf{m}_x \times \mathbf{m}_y) dS$. For finite $R$, $\rho_t$ is a finite real number, and for $R \to \infty$, $\rho_t \to 0$. Nevertheless, this transition does not change the topological charge of the whole space.

**Preliminary analysis of emergent elasticity and topological stability in reciprocal space**

Topological phase transitions are intrinsically local phenomena, but if we replace the real space by the reciprocal space, a local point now represents one or several global dynamical modes, which enables studies of topological changes of global properties. Now we apply the theory developed above to study the topological stability of the dynamical property of crystals. In this case, the state variable to be studied is the dispersion relation, and the topological stability is loss when the positive-definiteness of some branch of the dispersion relation is broken at certain point in the reciprocal space. We study three prototype examples: the phonon spectrum of a monoatomic chain and a biatomic chain under Born-Karman periodic boundary condition, and the band structure of a monoatomic chain within the Kronig-Penny model.

*A. Phonon spectrum of a monoatomic chain*

The Lagrangian of a monoatomic chain composed of $N$ identical atoms reads[19]

$$L = T - U = \frac{1}{2}m\sum_{n=1}^{N} \dot{u}_n^2 + \frac{1}{2}\beta\sum_{n=1}^{N}(u_n - u_{n-1})^2, \qquad (20)$$

where $T$ is the kinetic energy, $V$ is the potential energy, $m$ is the mass of the atom, $\beta$ is the force constant, and $u_n$ is the displacement of the $n^{\text{th}}$ atom. Substituting the lattice wave solution

$u_n = \frac{1}{\sqrt{Nm}} \sum_q Q(q,t) e^{-inaq}$, where $a$ is the lattice constant, at $N \to \infty$ eq. (20) can be rewritten as

$$L = T - U = \frac{1}{2}\int_{-\frac{\pi}{a}}^{\frac{\pi}{a}} |\dot{Q}(q,t)|^2 dq - \frac{1}{2}\int_{-\frac{\pi}{a}}^{\frac{\pi}{a}} \frac{2\beta}{m}(1 - \cos aq)|Q(q,t)|^2 dq, \quad (21)$$

where $q$ is the coordinate of the 1D reciprocal space. $Q_0(q,t) = A e^{i\omega(q)t}$ with $\omega^2(q) = \frac{2\beta}{m}(1 - \cos aq)$ is the stable solution of the system with a Lagrangian defined by eq. (21), where the stability is determined by the positive-definiteness of the potential energy $U$ [21], which gives $\frac{\delta^2 \phi}{\delta |Q|^2} = \omega^2(q) > 0 \ \forall\ q \in (-\frac{\pi}{a}, 0) \cup (0, \frac{\pi}{a}]$, where $\phi(q, Q(q,t))$ denotes the potential energy density ($U = \int_{-\frac{\pi}{a}}^{\frac{\pi}{a}} \phi dq$). The physical significance of this condition is that no soft modes exists in the reciprocal space except the $\Gamma$ point. In this case, we only have to analyze the domain emergent elasticity problem with emergent elastic energy density $\phi_u = \phi(q, Q_0(q - u(q), t))$ and a vanishing solution $u(q) = 0$. Since $\phi$ contains no gradient terms of $Q$, the topological stability of $Q_0(q,t)$ is determined by

$$\frac{\delta^2 \phi_u}{\delta u^2} = |A|^2 \omega'^2(q) \omega^2(q) t^2. \quad (22)$$

Since $\frac{\delta^2 \phi}{\delta |Q|^2} = \omega^2(q) > 0$, a critical condition of topological stability valid at any time can be derived from eq. (22) as

$$\omega'(q) = 0. \quad (23)$$

For monoatomic chain eq. (23) is satisfied only at $q = \frac{\pi}{a}$, where $\omega(q)$ reaches its maximum. That is to say, $\omega > \omega(\frac{\pi}{a})$ is the forbidden band of the system, and at the boundary of this forbidden band the emergent stiffness of $Q_0(q,t)$ is softened.

*B. Phonon spectrum of a diatomic chain*

The above analysis of the emergent stiffness and topological stability of phonon spectrum has fundamental significance. Within the harmonic approximation, the lattice dynamics of a crystalline structure with arbitrary complexity can be reduced to a set of independent phononic branches with their own dispersion relations. And for each of these dispersion relations, a critical condition of topological stability similar to eq. (23) can be derived. For biatomic chain we have

$$\begin{cases} \omega_a^2(q) = \beta \dfrac{1+b}{bm}\left(1 - \sqrt{1 - \dfrac{4b}{(1+b)^2}\sin^2 aq}\right), \\ \omega_o^2(q) = \beta \dfrac{1+b}{bm}\left(1 + \sqrt{1 - \dfrac{4b}{(1+b)^2}\sin^2 aq}\right), \end{cases} \quad (24)$$

where $\omega_a(q)$ denotes the dispersion relation of the acoustic branch, $\omega_o(q)$ denotes the dispersion relation of the optical branch, $m$ is the mass of the light atom, $b$ is the mass ratio between the heavy and light atom ($b \geq 1$). For convenience, by setting $\beta = m = a = 1$ (Fig. 5(a)), we have from eq. (24)

$$\begin{cases} \omega'_a(q) = \dfrac{\sin 2q}{f(q)\sqrt{1 + (1 - f(q))/b}}, \\ \omega'_o(q) = \dfrac{\sin 2q}{f(q)\sqrt{1 + (1 + f(q))/b}}, \end{cases} \quad (25)$$

where $f(q) = \sqrt{1 + b^2 + 2b\cos 2q}$. We study the distribution of $\omega'_a(q)$ and $\omega'_o(q)$ in the Brillouin zone at different values of $b$ (Fig. 5(b)). At $b = 1$, a biatomic chain reduces to a monoatomic chain, where the band gap between the acoustic branch and the optical branch is closed. The close and open of a band gap obviously changes the topological property of the dispersion relations, which is clearly explained by the critical condition eq. (23) using eq. (25). At $b = 1$, we have $\omega'_a(\frac{\pi}{2}) = -\omega'_o(\frac{\pi}{2}) = \frac{\sqrt{2}}{2}$. once $b$ increases from 1, $\omega'_a(\frac{\pi}{2})$ and $\omega'_o(\frac{\pi}{2})$ reduce sharply to zero (Figure 2), indicating that band gap opening is a second-order TPT initiated by softened of the emergent stiffness at $q = \frac{\pi}{2}$ in the reciprocal space.

*C. Band structure of a monoatomic chain within the Kronig-Penny model*

The above analysis of the emergent stiffness and topological stability in reciprocal space is applicable to classic systems as well as quantum systems. We study the band structure of a monoatomic chain within the Kronig-Penny model, whose band dispersion relation can be determined by [20]

$$(P/Ka)\sin Ka + \cos Ka = \cos qa, \quad (26)$$

where $\epsilon = \hbar^2 K^2/2m$ is the eigenvalue of energy, $P$ is a constant representing the strength of periodic potential, a is the lattice constant. $K = K_0(q)$ can be numerically determined from eq. (17), and with $\epsilon'(q) = \dfrac{\hbar^2}{m}\left(K\dfrac{dK}{dq}\right)_0$ one can finally obtain the variation of $\epsilon'(q)$ with $q$. Setting

$P = \frac{3\pi}{2}$, $a = 1$, we show in Fig. 6 the dispersion relation as well as the variation of $\epsilon'(q)$ with $q$ in $q \in (0, 4\pi)$. It is shown that on both sides of the boundary of a band gap, we have $\epsilon'(q) = 0$. Compare the band structure of a free electron with that in Fig. 6(a), we can say that introduction of a periodic potential in a monoatomic chain leads to the band-gap-openning, which is a typical second-order TPT, reflected by softening of $\epsilon'(q)$ at those points of $q$ where the band gap opens.

**Acknowledgments**

The work was supported by the NSFC (National Natural Science Foundation of China) through the funds with Grant Nos. 12172090, 11772360, 11832019.

**Author contributions**

Y.H. conceived the idea and conducted the work.

**Competing interests**

The authors declare no competing interests.

**References**

*[1] Landau, L. D. (1937). On the theory of phase transitions. I. Zh. Eksp. Teor. Fiz., 11, 19.*

*[2] Kosterlitz, J. M. and Thouless, D. J. (1972). Long range order and metastability in two-dimensional solids and superfluids. J. Phys. C: Solid State Phys. 5 L124–6.*

*[3] Kosterlitz, J. M. and Thouless, D. J. (1973). Ordering, metastability and phase transitions in two-dimensional systems. J. Phys. C: Solid State Phys. 6 1181–203.*

*[4] Tsui, D. C., H. L. Stormer, and A. C. Gossard, (1982). Two-dimensional magnetotransport in the extreme quantum limit. Phys. Rev. Lett. 48, 1559.*

[5] Wen, X. G. (1990). *Topological orders in rigid states*. International Journal of Modern Physics B, 4(02), 239-271.

[6] Wen, X. G. (1995). *Topological orders and edge excitations in fractional quantum Hall states*. Advances in Physics, 44(5), 405-473.

[7] Wen, X. G. (2017). *Colloquium: Zoo of quantum-topological phases of matter*. Reviews of Modern Physics, 89(4), 041004.

[8] Toledano, P., & Toledano, J. C. (1987). *Landau Theory Of Phase Transitions, The Application To Structural, Incommensurate, Magnetic And Liquid Crystal Systems* (Vol. 3). World Scientific Publishing Company.

[9] Cyrot, M. (1973). *Ginzburg-Landau theory for superconductors*. Reports on Progress in Physics, 36(2), 103.

[10] Prigogine, I., & Nicolis, G. (1967). *On symmetry-breaking instabilities in dissipative systems*. The Journal of Chemical Physics, 46(9), 3542-3550.

[11] Prigogine, I., & Lefever, R. (1968). *Symmetry breaking instabilities in dissipative systems. II*. The Journal of Chemical Physics, 48(4), 1695-1700.

[12] Prigogine, I., Lefever, R., Goldbeter, A., & Herschkowitz-Kaufman, M. (1969). *Symmetry breaking instabilities in biological systems*. Nature, 223(5209), 913-916.

[13] Gelfand, I. M., & Silverman, R. A. *Calculus of variations*. (Courier Corporation, 2000).

[14] Weinberg, E. *Classical Solutions in Quantum Field Theory: Solitons and Instantons in High Energy Physics* (Cambridge Monographs on Mathematical Physics, 2012).

[15] Manton, N., & Sutcliffe, P. (2004). *Topological Solitons* (Cambridge Monographs on Mathematical Physics). Cambridge: Cambridge University Press. doi:10.1017/CBO9780511617034

[16] Perring, J. K., & Skyrme, T. H. R. (1962). *A model unified field equation*. Nucl. Phys., 31, 550-555.

[17] Roessler, U. K., Bogdanov, A. N., & Pfleiderer, C. (2006). *Spontaneous skyrmion ground states in magnetic metals*. Nature, 442(7104), 797-801.

[18] Mühlbauer, S., Binz, B., Jonietz, F., Pfleiderer, C., Rosch, A., Neubauer, A., ... & Böni, P. (2009). *Skyrmion lattice in a chiral magnet*. Science, 323(5916), 915-919.

[19] Born, M. & Huang, K. *Dynamical Theory of Crystal Lattices*. (Clarendon, Oxford, 1954).

[20] Kittel, C., McEuen, P., & McEuen, P. Introduction to solid state physics (New York: Wiley, 1996).

[21] Berdichevsky, V. L. Variational Principles of Continuum Mechanics: I. Fundamentals. (Springer Science & Business Media, 2009).

[22] Hu, Y. (2022) Emergent elasticity and topological stability of solitons. Submitted.

[23] Altland, A., & Simons, B. (2010). Condensed Matter Field Theory (2nd ed.). Cambridge: Cambridge University Press. doi:10.1017/CBO9780511789984.

[24] Leonov, A. O., & Bogdanov, A. N. (2018). Crossover of skyrmion and helical modulations in noncentrosymmetric ferromagnets. New Journal of Physics, 20(4), 043017.

[25] Leonov, A. O., Dragunov, I. E., Rößler, U. K., & Bogdanov, A. N. (2014). Theory of skyrmion states in liquid crystals. Physical Review E, 90(4), 042502.

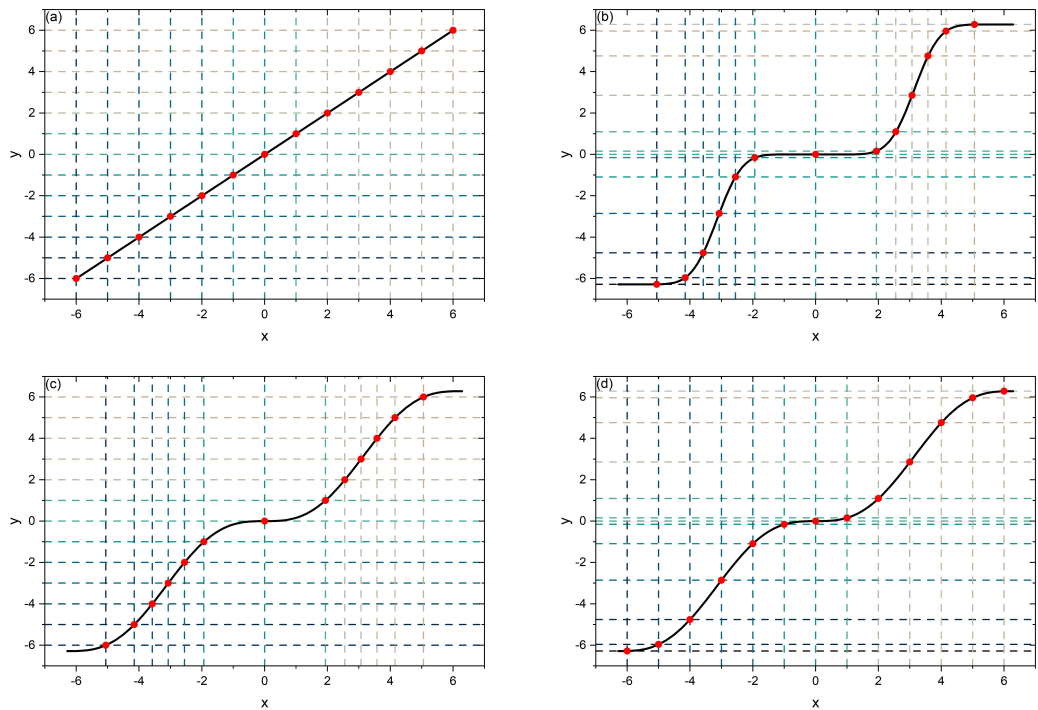

Figure 1. Image of the function $(y + v(y))(x - u(x))$, where (a) $y(x) = x, u(x) = 0, v(y) = 0$, (b) $y(x) = x, u(x) = \sin x, v(y) = \sin y$, (c) $y(x) = x, u(x) = \sin x, v(y) = 0$, and (d) $y(x) = x, u(x) = 0, v(y) = \sin y$.

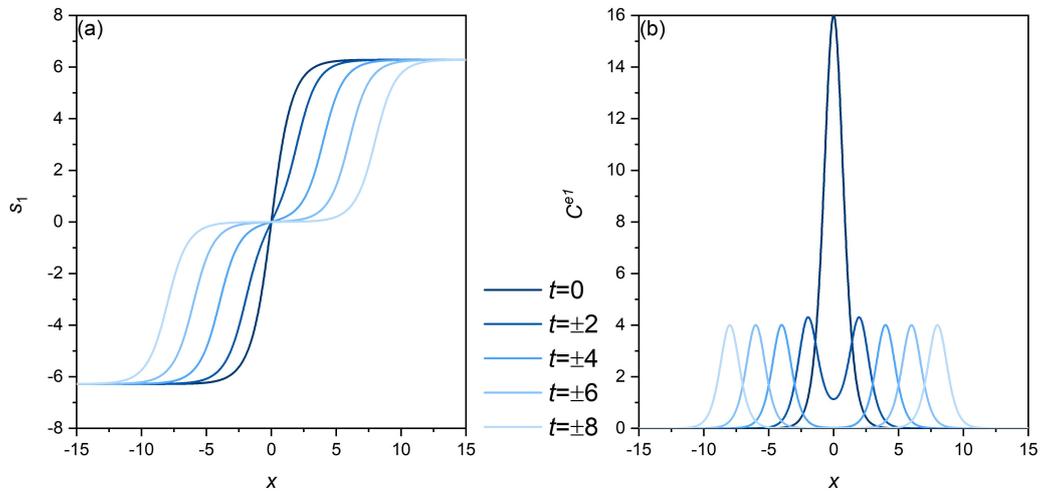

Figure 2. (a) Field pattern and (b) emergent stiffness $C^{e1}$ of the kink-kink collision solution of the sine-Gordon model at different values of $t$.

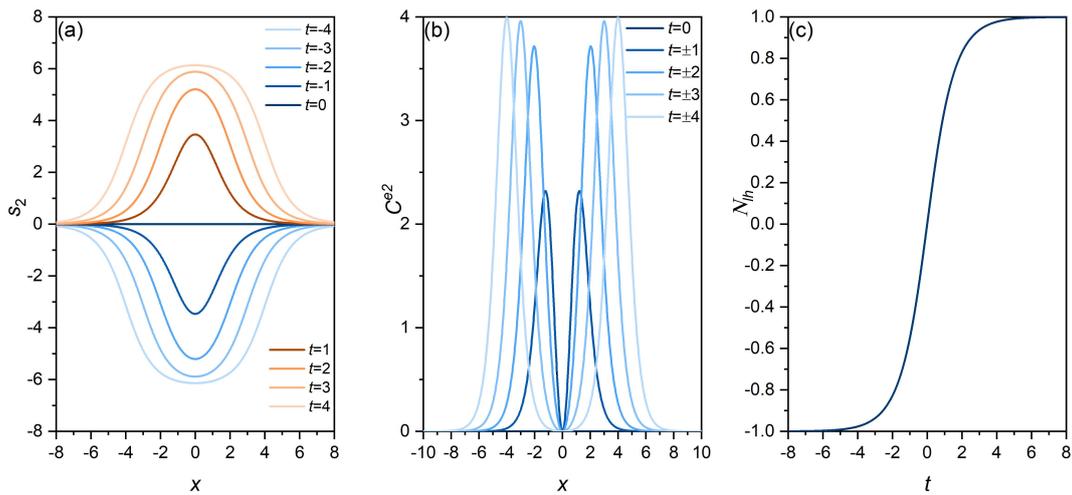

Figure 3. (a) Field pattern and (b) emergent stiffness $C^{e2}$ of the kink-antikink collision solution of the sine-Gordon model at different values of $t$. (c) Variation in the topological charge of the left half-space with $t$.

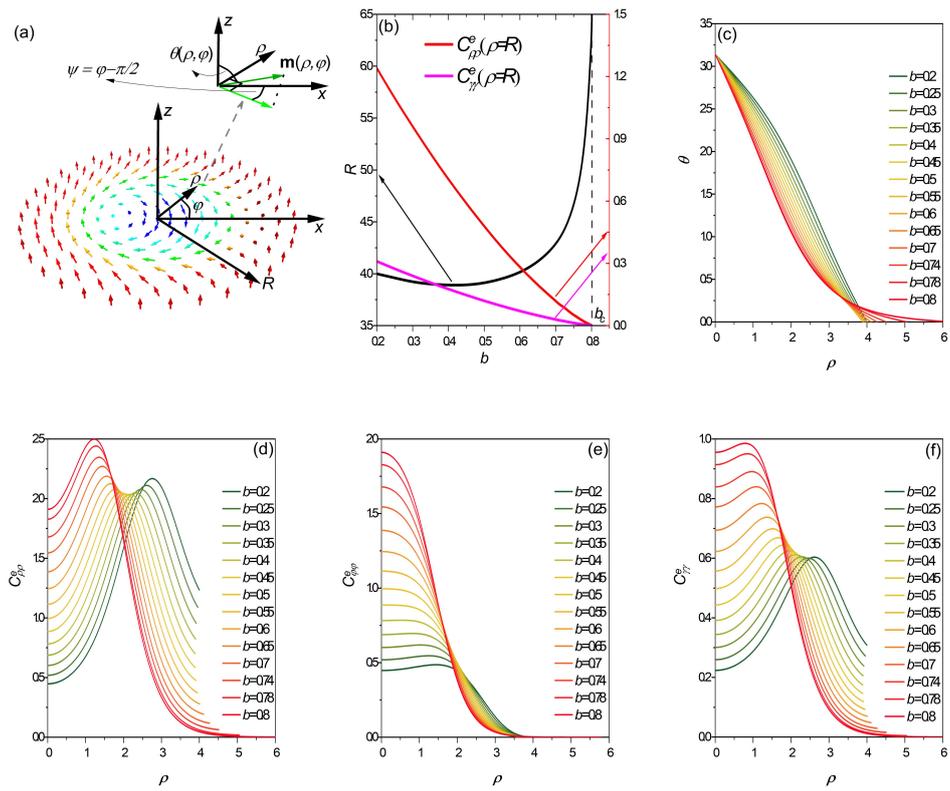

Figure 4. (a) Illustration of the field pattern of a Bloch-type isolated magnetic skyrmion. (b) Variation in $R$, $C^e_{\rho\rho}(\rho = R)$ and $C^e_{\gamma\gamma}(\rho = R)$ with $b$. Variation in (c) $\theta$, (d) $C^e_{\rho\rho}$, (e) $C^e_{\varphi\varphi}$, and (f) $C^e_{\gamma\gamma}$ with $\rho$ at different values of $b$.

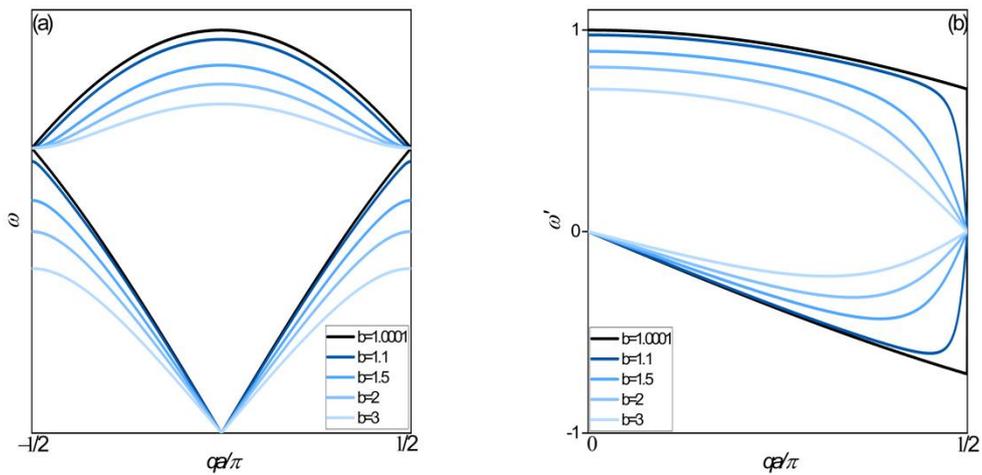

Figure 5. (a) Dispersion relation and (b) variation of $\frac{d\omega}{dq}$ with $q$ of a diatomic chain with different parameters of mass ratio $b$.

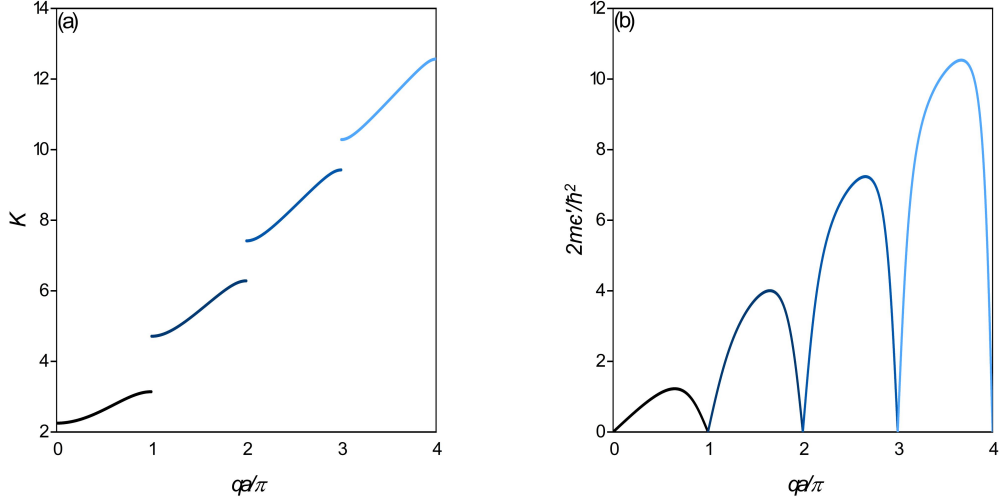

Figure 6. (a) Band structure and (b) variation of $\frac{2m}{\hbar^2}\frac{d\epsilon}{dq}$ with $q$ of a mono-atomic chain with different parameters of mass ratio $b$.

## Methods

*A. General vanishing solution of the emergent elasticity problem*

We first consider the domain emergent elasticity problem of the stable solution $y = y_0(x)$ of the Euler-Lagrangian equation $\frac{d}{dx}\left(\frac{\partial \phi}{\partial y'}\right) - \frac{\partial \phi}{\partial y} = 0$. For $y_0(x) \to y_0(x - u(x))$, we have $\phi_u = \phi(x, y_0(x - u(x)), \frac{d}{dx}y_0(x - u(x)))$, which gives after manipulation

$$\delta\phi_u = \left[\frac{\partial \phi}{\partial y} - \frac{d}{dx}\left(\frac{\partial \phi}{\partial y'}\right)\right]_u y_0'\delta u, \tag{A1}$$

where a subscript $u$ means the term takes value at $y = y_0(x - u(x))$. Therefore $\frac{\delta\phi_u}{\delta u} = 0$ requires that $\left[\frac{\partial \phi}{\partial y} - \frac{d}{dx}\left(\frac{\partial \phi}{\partial y'}\right)\right]_u = 0$ or $y_0'(x) = 0$ for any $x$. Since $y_0(x)$ can be any stable solution, $y_0'(x) = 0$ cannot be guaranteed, which gives $\left[\frac{\partial \phi}{\partial y} - \frac{d}{dx}\left(\frac{\partial \phi}{\partial y'}\right)\right]_u = 0$. Since $u(x)$ is small everywhere, and we have $\left[\frac{\partial \phi}{\partial y} - \frac{d}{dx}\left(\frac{\partial \phi}{\partial y'}\right)\right]_0 = 0$, $u(x) = 0$ is obviously the solution of $\left[\frac{\partial \phi}{\partial y} - \frac{d}{dx}\left(\frac{\partial \phi}{\partial y'}\right)\right]_u = 0$.

We then consider the range emergent elasticity problem of the stable solution $y = y_0(x)$ of the Euler-Lagrangian equation $\frac{d}{dx}\left(\frac{\partial \phi}{\partial y'}\right) - \frac{\partial \phi}{\partial y} = 0$. For $y_0(x) \to y_0(x) + v(y_0(x))$, we have $\phi_v = \phi(x, y_0(x) + v(y_0(x)), \frac{d}{dx}[y_0(x) + v(y_0(x))])$, which gives after manipulation

$$\delta\phi_v = \left[\frac{\partial\phi}{\partial y} - \frac{d}{dx}\left(\frac{\partial\phi}{\partial y'}\right)\right]_v \delta v, \tag{A2}$$

where a subscript $v$ means the term takes value at $y = y_0(x) + v(y_0(x))$. Therefore $\frac{\delta\phi_u}{\delta v} = 0$ requires that $\left[\frac{\partial\phi}{\partial y} - \frac{d}{dx}\left(\frac{\partial\phi}{\partial y'}\right)\right]_v = 0$. Since $v(y)$ is small everywhere, and we have $\left[\frac{\partial\phi}{\partial y} - \frac{d}{dx}\left(\frac{\partial\phi}{\partial y'}\right)\right]_0 = 0$, $v(y_0) = 0$ is obviously the solution of $\left[\frac{\partial\phi}{\partial y} - \frac{d}{dx}\left(\frac{\partial\phi}{\partial y'}\right)\right]_v = 0$.

*B. Equivalence of domain emergent elastic stiffness tensor and range emergent elastic stiffness tensor for some regular form of free energy density*

Without loss of generality, assume that $m = n = 3$, for a vector field solution $\mathbf{p}(\mathbf{x}) = \tilde{\mathbf{p}}(\mathbf{x})$ of a system with a free energy density $\phi = \phi(\mathbf{x}, p_i, p_{j,k})$, we do the following replacements concerning domain emergent elasticity: $p_i(\mathbf{x}) \to \tilde{p}_i(\mathbf{x} - \mathbf{u}(\mathbf{x}))$, $p_{j,k}(\mathbf{x}) \to \tilde{p}_{j,i}(\mathbf{x} - \mathbf{u}(\mathbf{x}))(\delta_{ik} - u_{i,k})$, and obtain $\phi_u = \phi(\mathbf{x}, \tilde{p}_i(\mathbf{x} - \mathbf{u}(\mathbf{x})), \tilde{p}_{j,i}(\mathbf{x} - \mathbf{u}(\mathbf{x}))(\delta_{ik} - u_{i,k}))$. In this case, we can define the following domain emergent elastic deformation vector

$$\boldsymbol{\varepsilon} = [u_{1,1} \quad u_{1,2} \quad u_{1,3} \quad u_{2,1} \quad u_{2,2} \quad u_{2,3} \quad u_{3,1} \quad u_{3,2} \quad u_{3,3}]^T, \tag{B1}$$

and the the domain emergent elastic stiffness matrix $\mathbf{C}^\varepsilon$ is derived by

$$C^\varepsilon_{ij} = \left(\frac{\partial^2 \phi_u}{\partial \varepsilon_i \partial \varepsilon_j}\right)_{\mathbf{u}(\mathbf{x})=\mathbf{0}}. \tag{B2}$$

Or we can define an equivalent fourth order tensor as follow

$$C^\varepsilon_{ijkl} = \left(\frac{\partial^2 \phi_u}{\partial u_{i,j} \partial u_{k,l}}\right)_{\mathbf{u}(\mathbf{x})=\mathbf{0}}, \tag{B3}$$

which can be further explicitly derived as

$$C^\varepsilon_{ijkl} = \left(\frac{\partial^2 \phi}{\partial p_{m,j} \partial p_{n,l}} p_{m,i} p_{n,k}\right)_{\mathbf{p}(\mathbf{x})=\tilde{\mathbf{p}}(\mathbf{x})}. \tag{B4}$$

Similarly, we do the following replacements concerning range emergent elasticity: $p_i(\mathbf{x}) \to \tilde{p}_i(\mathbf{x}) + v_i(\tilde{\mathbf{p}}(\mathbf{x}))$, $p_{j,k}(\mathbf{x}) \to \tilde{p}_{j,k}(\mathbf{x}) + v_{j;i}(\tilde{\mathbf{p}}(\mathbf{x}))\tilde{p}_{i,k}(\mathbf{x})$, where $v_{j;i}(\tilde{\mathbf{p}}(\mathbf{x})) = \left(\frac{\partial v_j(\mathbf{p}(\mathbf{x}))}{\partial p_i(\mathbf{x})}\right)_{\mathbf{p}(\mathbf{x})=\tilde{\mathbf{p}}(\mathbf{x})}$ and obtain $\phi_u = \phi(\mathbf{x}, \tilde{p}_i(\mathbf{x}) + v_i(\tilde{\mathbf{p}}(\mathbf{x})), \tilde{p}_{j,k}(\mathbf{x}) + v_{j;i}(\tilde{\mathbf{p}}(\mathbf{x}))\tilde{p}_{i,k}(\mathbf{x}))$. In this case, we can define the following range emergent elastic deformation vector

$$\mathbf{w} = [v_{1;1} \quad v_{1;2} \quad v_{1;3} \quad v_{2;1} \quad v_{2;2} \quad v_{2;3} \quad v_{3;1} \quad v_{3;2} \quad v_{3;3}]^T, \tag{B5}$$

and the the range emergent elastic stiffness matrix $\mathbf{C}^\mathbf{w}$ is derived by

$$C^w_{ij} = \left(\frac{\partial^2 \phi_u}{\partial w_i \partial w_j}\right)_{\mathbf{v}(\tilde{\mathbf{p}}(\mathbf{x}))=\mathbf{0}}. \tag{B6}$$

Or we can define an equivalent fourth order tensor as follow

$$C^w_{ijkl} = \left(\frac{\partial^2 \phi_u}{\partial v_{i;j} \partial v_{k;l}}\right)_{\mathbf{v}(\tilde{\mathbf{p}}(\mathbf{x}))=\mathbf{0}}, \tag{B7}$$

which can be further explicitly derived as

$$C^w_{ijkl} = \left(\frac{\partial^2 \phi}{\partial p_{m,j} \partial p_{n,l}} p_{m,i} p_{n,k}\right)_{\mathbf{p}(\mathbf{x})=\tilde{\mathbf{p}}(\mathbf{x})}. \tag{B8}$$

Compare eq. (B4) and eq. (B8) we see immediately that $\mathbf{C}^\varepsilon = \mathbf{C}^w$ for any system whose free energy density takes the form $\phi = \phi(\mathbf{x}, p_i, p_{j,k})$.

*C. Emergent elasticity and topological stability of a vector field in 3D space*

For an arbitrary vector solution $\mathbf{p}(\mathbf{x}) = \tilde{\mathbf{p}}(\mathbf{x})$ with three components in 3D space, a general form of critical condition of topological stability can be derived and is presented as follow. For $\phi = \phi(\mathbf{x}, p_i, p_{j,k})$, we do the following replacements concerning domain emergent elasticity: $p_i(\mathbf{x}) \rightarrow \tilde{p}_i(\mathbf{x} - \mathbf{u}(\mathbf{x}))$, $p_{j,k}(\mathbf{x}) \rightarrow \tilde{p}_{j,i}(\mathbf{x} - \mathbf{u}(\mathbf{x}))(\delta_{ik} - u_{i,k}) = \tilde{p}_{j,i}(\mathbf{x} - \mathbf{u}(\mathbf{x}))(\delta_{ik} - (\varepsilon_{ik} + \epsilon_{ikl}\omega_l))$, and obtain $\phi_u = \phi(\mathbf{x}, \tilde{p}_i(\mathbf{x} - \mathbf{u}(\mathbf{x})), \tilde{p}_{j,i}(\mathbf{x} - \mathbf{u}(\mathbf{x}))(\delta_{ik} - (\varepsilon_{ik} + \epsilon_{ikl}\omega_l)))$, where $\varepsilon_{ik} = \frac{1}{2}(u_{i,k} + u_{k,i})$ is the domain emergent elastic strains and $\epsilon_{ikl}\omega_l = \frac{1}{2}(u_{i,k} - u_{k,i})$, with $\omega_l$ the emergent rotational angle and $\epsilon_{ikl}$ are components of the Levi-Civita tensor according to relevant definition in solid mechanics. In 3D space, we can define the following domain emergent elastic deformation vector

$$\boldsymbol{\varepsilon} = [\varepsilon_{11} \quad \varepsilon_{22} \quad \varepsilon_{33} \quad \varepsilon_{23} \quad \varepsilon_{13} \quad \varepsilon_{12} \quad \omega_1 \quad \omega_2 \quad \omega_3]^T, \tag{C1}$$

and the domain emergent elastic stiffness matrix $\mathbf{C}^\varepsilon$ is derived by

$$C^\varepsilon_{ij} = \left(\frac{\partial^2 \phi_u}{\partial \varepsilon_i \partial \varepsilon_j}\right)_{\mathbf{u}(\mathbf{x})=\mathbf{0}}. \tag{C2}$$

The critical condition of topological stability in this case is at some point or points in space, at least one of the eigenvalues of $\mathbf{C}^\varepsilon$ drops to zero. According to Methods B, in this case the range emergent elastic stiffness matrix is equivalent to $\mathbf{C}^\varepsilon$, so that $\mathbf{C}^\varepsilon$ can be referred to as the emergent elastic stiffness matrix.

For $\phi = \phi(\mathbf{x}, p_i)$, we do the following replacements concerning domain emergent elasticity: $p_i(\mathbf{x}) \rightarrow \tilde{p}_i(\mathbf{x} - \mathbf{u}(\mathbf{x}))$, and obtain $\phi_u = \phi(\mathbf{x}, \tilde{p}_i(\mathbf{x} - \mathbf{u}(\mathbf{x})))$, the state variable here is the emergent displacement vector

$$\mathbf{u} = [u_1 \quad u_2 \quad u_3]^T, \tag{C3}$$

and the the domain emergent displacement stiffness matrix $\mathbf{C}^u$ is derived by

$$C^u_{ij} = \left(\frac{\partial^2 \phi_u}{\partial u_i \partial u_j}\right)_{\mathbf{u}(\mathbf{x})=\mathbf{0}}. \tag{C4}$$

The critical condition of topological stability in this case is at some point or points in space, at least one of the eigenvalues of $\mathbf{C^u}$ drops to zero. For range emergent elasticity of this case, a condition similar to that of the second order variation of $\phi$ with respect to $p_i$ can be obtained, so that $\mathbf{C^u}$ can be referred to as the emergent displacement stiffness matrix.

Now we apply the general formula above to study the isolated magnetic skyrmion. The free energy density is given by eq. (18), which corresponds to the case $\phi = \phi(m_i, m_{j,k})$, and the soliton solution is given by eq. (19), which takes the form $\mathbf{m} = \mathbf{m}(\rho) = \mathbf{m}(\sqrt{x^2+y^2})$. Since this is a 2D solution, eq. (C1) reduces to

$$\boldsymbol{\varepsilon} = [\varepsilon_{11} \quad \varepsilon_{22} \quad \varepsilon_{12} \quad \omega_3]^T. \tag{C5}$$

And for the free energy density given by eq. (18), to obtain the expression of $\phi_u$ we do the following replacements: $m_{1,1} \to (1-\varepsilon_{11})m_{1,1} - (\varepsilon_{12} - \omega_3)m_{1,2}, m_{1,2} \to -(\varepsilon_{12} + \omega_3)m_{1,1} + (1-\varepsilon_{22})m_{1,2}, m_{2,1} \to (1-\varepsilon_{11})m_{2,1} - (\varepsilon_{12} - \omega_3)m_{2,2}, m_{2,2} \to -(\varepsilon_{12} + \omega_3)m_{2,1} + (1-\varepsilon_{22})m_{2,2}, m_{3,1} \to (1-\varepsilon_{11})m_{3,1} - (\varepsilon_{12} - \omega_3)m_{3,2}, m_{3,2} \to -(\varepsilon_{12} + \omega_3)m_{3,1} + (1-\varepsilon_{22})m_{3,2}$, and from eq. (C2) the expressions of the emergent elastic stiffness matrix $\mathbf{C^\varepsilon}$ can be derived in the Cartesian coordinates. Since the solution of isolated magnetic skyrmion in this case is axially symmetric, it is more convenient to study its emergent elastic stiffness in polar coordinates. To achieve this, we recall the following coordinate transformation relation:

$$\hat{\boldsymbol{\varepsilon}} = \mathbf{K}\boldsymbol{\varepsilon}, \tag{C6}$$

where

$$\mathbf{K} = \begin{bmatrix} \cos^2\varphi & \sin^2\varphi & -\sin 2\varphi/2 & 0 \\ \sin^2\varphi & \cos^2\varphi & \sin 2\varphi/2 & 0 \\ \sin 2\varphi & -\sin 2\varphi & \cos 2\varphi & 0 \\ 0 & 0 & 0 & 1 \end{bmatrix}, \tag{C7}$$

$$\hat{\boldsymbol{\varepsilon}} = [\varepsilon_{\rho\rho} \quad \varepsilon_{\varphi\varphi} \quad \varepsilon_{\rho\varphi} \quad \omega_3]^T. \tag{C8}$$

The emergent elastic stiffness matrix presented in polar coordinates can be derived as

$$\widehat{\mathbf{C}}^\varepsilon = \mathbf{K}^T \mathbf{C}^\varepsilon \mathbf{K}. \tag{C9}$$

When $\phi$ is given by eq. (18), we have after manipulation

$$\widehat{\mathbf{C}}^\varepsilon = \begin{bmatrix} C^e_{\rho\rho} & 0 & 0 & 0 \\ 0 & C^e_{\varphi\varphi} & 0 & 0 \\ 0 & 0 & C^e_{\gamma\gamma} & 0 \\ 0 & 0 & 0 & C^{\omega_3} \end{bmatrix}, \tag{C10}$$

where $C^e_{\rho\rho} = 2\theta'^2(\rho)$, $C^e_{\varphi\varphi} = \frac{2\sin^2\theta(\rho)}{\rho^2}$, $C^e_{\gamma\gamma} = 2\left(\theta'^2(\rho) + \frac{\sin^2\theta(\rho)}{\rho^2}\right)$, and $C^{\omega_3} = \frac{2\sin^2\theta(\rho)}{\rho^2} + 2\theta'^2(\rho)$. Since $\widehat{\mathbf{C}}^\varepsilon$ is already diagonal, $C^e_{\rho\rho}$, $C^e_{\varphi\varphi}$, $C^e_{\gamma\gamma}$ and $C^{\omega_3}$ are four eigenvalues of $\widehat{\mathbf{C}}^\varepsilon$ and the corresponding eigenvectors are $[1 \ 0 \ 0 \ 0]^T$, $[0 \ 1 \ 0 \ 0]^T$, $[0 \ 0 \ 1 \ 0]^T$, and $[0 \ 0 \ 0 \ 1]^T$, respectively.